\newcommand\arcsec{\mbox{$.\!^{\prime\prime}$}}%
\newcommand\fm{\mbox{$.\!\!^{\mathrm m}$}}%
\newcommand\kms{\nobreak\mbox{$\;$km\,s$^{-1}$}}

\newcommand\la{\mathrel{\hbox{\rlap{\hbox{\lower4pt\hbox{$\sim$}}}\hbox{$<$}}}}
\newcommand\ga{\mathrel{\hbox{\rlap{\hbox{\lower4pt\hbox{$\sim$}}}\hbox{$>$}}}}

\documentclass[11pt]{article}
\usepackage{moriond,epsfig}

\usepackage{natbib}
\bibpunct{(}{)}{;}{a}{}{,}

\begin{document}
\vspace*{4cm}
\title{COSMIC EXPANSION AND H\boldmath{$_0$}: \\ 
       A RETRO- AND PRO-SPECTIVE NOTE}

\author{ G.A. TAMMANN \& B. REINDL }

\address{Astronomisches Institut der Universit\"at Basel \\
Venusstrasse 7, CH-4102 Binningen, Switzerland}

\pagestyle{plain}
\maketitle
\abstracts{
   Attempts to measure extragalactic distances over the last 90 years
   are briefly described. It follows a short history of the discovery
   of the expansion of space. Reasons are discussed for the decrease
   of the Hubble constant from $H_{0}\approx500$ originally to
   $H_{0}\la60$ at present. Remaining problems with Cepheids as local
   distance calibrators are outlined.
}

\section{Introduction}
The conquest of the third dimension beyond the Galaxy was one of the
great challenges of the 20th century. It is not possible to do justice
to its fascinating and complex history in a few pages. Much of the
progress during the first half of the century was made at Mount
Wilson. Insight into this part of the history is provided by
\citet{Sandage:95}, and his forthcoming monograph ``An informal
history of the Mount Wilson Observatory (1904$-$1950)'' will be the
prime reference for the subject with all its ramifications into most
fields of astronomy.

\section{Early Galaxy Distances}
The battle about the nature of the ``nebulae'' was still raging
\citep[cf.][]{Fernie:70} when the first attempts were made to measure
the distances of individual galaxies. Five steps should particularly
be remembered: 
\begin{itemize}
\item Hertzsprung's (\citeyear{Hertzsprung:14}) tragedy. He had
  determined the distance modulus of SMC from Cepheids to be
  $(m-M)_{\rm SMC}=20.3$ (115\,kpc), roughly a factor 1.8 too
  large. According to the custom of the time he transformed the
  distance into a trigonometric parallax of $0\arcsec0001$, losing a
  factor of 10 during the process. While transforming the 
  parallax into light years he lost another factor of ten. Thus his
  published distance of 3000 light years buried his sensational
  result.
\item Shapley's (\citeyear{Shapley:15,Shapley:18}) mistake. In the
  following year Shapley repeated Hertzsprung's measurement. For
  various reasons he now obtained a Cepheid distance of only
  $(m-M)_{\rm SMC}=16.1$ (17\,kpc), which he slightly increased in
  1918 and which he could take as a confirmation of his conviction
  that all ``nebulae'' were part of his very large Galactic system.
\item Lundmark's (\citeyear{Lundmark:20}) insight: He was the first
  to recognize supernovae as a class distinct from novae. This
  explained the brightness of the ``nova'' 1885 in M\,31 and led him
  to a modulus of $(m-M)_{\rm M31}=21.3$ (180\,kpc). Still much too
  low the value could not be accommodated within even the wildest size
  estimates of the Galaxy. But the result had no influence on the
  ``Great Debate''.
\item {\"O}pik's (\citeyear{Oepik:21,Oepik:22}) ingenuity. He used
  the rotation velocity of M\,31 to determine the mass-to-light ratio
  of the galaxy and he broke the distance degeneracy of this value by
  adopting a very reasonable mass-to-light ratio of the Solar
  neighborhood. He obtained a stunningly good value of $(m-M)_{\rm
    M31}=24.5$ (750\,kpc), which he decreased in the following year
  by a factor of 1.7. {\"O}pik's papers remained unnoticed.
\item Hubble's breakthrough in 1924. The discovery of several novae in
  ``nebulae'', first by \citet{Ritchey:17}, stimulated the search for
  variability and led Hubble to the discovery of a Cepheid in M\,31 in
  1923, --- the first Cepheid beyond the Magellanic Clouds. At the
  meeting of the Association for the Advancement of Science in
  December 1924 he announced the discovery of several very faint
  Cepheids in M\,31. They proved that many of the nebulae are actually
  ``island universes'', but the proof was not generally accepted for a
  while, because van Maanen's (\citeyear{vanMaanen:23}) unexplained
  measurement from proper motions of a detectable rotation of the
  spirals \citep[for a possible exlanation cf.][]{Baade:63} stood
  still in the way. Hubble published his Cepheid distance of M\,31
  only in \citeyear{Hubble:29a}, after he had published the Cepheids
  in NGC\,6822 (\citeyear{Hubble:25}) and M\,33
  (\citeyear{Hubble:26}).    
\end{itemize}

\section{The Emergence of Cosmic Expansion}
The great paradigms are often introduced on humble paths. Did
\citet{Slipher:14} realize that his first -- and subsequent -- radial
velocities of ``nebulae'' would be the stepping stone of modern
cosmology? The first interpretations of the surprisingly large radial
velocities were made under the impression of the
``\citet{deSitter:16,deSitter:17} effect'', which was a correct, yet
unphysical solution of Einstein's field equation, predicting a change
of the clock rate with the square of the distance $r$ in a {\em
  static\/} Universe
\citep{Wirtz:18,Wirtz:25,Stromberg:25,Lundmark:25}. Lundmark wrote ``the
smaller and presumably more distant spirals have a higher space
motion'', yet his ansatz $v=ar+br^2+c$ betrays his preoccupation with
the de Sitter effect.

   \citet{Lemaitre:27} was the first to combine observations with the
theory of expanding space. Extrapolating Hubble's Cepheid distances,
he derived a value of what was to become known as the ``Hubble
constant'' of $H_{0}=627\;$[km\,s$^{-1}$\,Mpc$^{-1}$]. A similar
analysis of the observations led \citet{Robertson:28} to imply a value
of $H_0=461$. 

   Hubble had used his Cepheid distances to calibrate the brightest
stars ($M_{\rm pg}=-6.3$) and the mean luminosity of bright galaxies
($M_{\rm pg}=-15.8$; --- either value $4^{\rm m}-5^{\rm m}$ too
faint). In this way he extended his distance scale out to the Virgo
cluster. \citet{Hubble:29b} plotted his distances against Slipher's
radial velocities and derived a value of $H_{0}=500$. This is
generally considered as the discovery of the expanding Universe. But
he was not sure what to make of his discovery and he commented:
``\dots the relation may represent the de Sitter [clock]
effect''. Considerable weight of Hubble's solution hinges on four
Virgo cluster members, which would nowadays be excluded because of
their virial motions. Hubble was lucky that Slipher had not
accidentally picked the bright Virgo spiral NGC\,4569 with its
negative radial velocity; this galaxy would have considerably confused
his first Hubble diagram. But he knew already of M. Humason's very
large velocity of NGC\,7619 (3800\kms), and in the following years he
extended the Hubble diagram in collaboration with Humason beyond
40\,000\kms \citep{Humason:36}. It is curious that Hubble, while
convincing the world of the expanding Universe with his brilliant book
``{\em The Realm of the Nebulae}'' (\citeyear{Hubble:36a}), died
with doubts on his mind about the expansion. They were based on his
unsatisfactory counts of galaxies (\citealt{Hubble:36b};
cf.\ \citealt{Sandage:95}), and his last Hubble diagram
(\citeyear{Hubble:53}) is commented ``no recession factor applied'',
i.e.\ he did not apply the full K-correction required in an expanding
Universe (cf.\ \citealt{Sandage:95}).

   \citet{deSitter:30} extended the Hubble diagram to $7800\kms$
including new redshifts by F.G Pease and M. Humason. Hubble was 
outraged and took it as an undue competition. de Sitter used distances
by Shapley, Lundmark, and Hubble as calibrators and galaxy diameters
as distance indicators. His data imply a Hubble constant just short of
500. Having discarded his clock rate effect he proved deep insight into
his result, still exceptional at the time, by writing: ``The only
acceptable explanation \dots is by the inherent expanding tendency of
space which follows from the solution \dots of Einstein's differential
equations for the inertia field''. 

   Any possibly remaining doubt about the expansion of space was
dispelled (it took a Fritz Zwicky to remain reluctant) by the fundamental
paper by \citet*[][known as HMS]{Humason:etal:56}. The authors
presented redshifts for 920 galaxies, objectively measured
magnitudes, the first correct K-corrections, and a Hubble diagram of
field galaxies as well as a tight Hubble diagram of clusters.

   In the following years Sandage (e.g.\ \citeyear{Sandage:72};
\citealt{Sandage:Hardy:73}; \citealt*{Sandage:etal:76}) started a vast
program to extend and to tighten the Hubble diagram of brightest
cluster members with the final goal to determine the deceleration
parameter $q_0$. The program came to a halt when it become clear that
(cluster) galaxies undergo luminosity evolution. But the results were
decisive subsequent to the discovery of the large quasar redshifts
which stimulated a mysticism, even among astronomers, which is now hard
to understand.

   We will return to the ultimate Hubble diagram of supernovae of type
Ia in Section~\ref{sec:08}, which does not only lead to the
large-scale value of $H_0$, but with considerable likelihood even to a
combined value of $\Omega_{\rm M}$ and $\Omega_{\Lambda}$.

\section{The Stretching of the Distance Scale}
\label{sec:04}
\citet{Hubble:36b} has decreased his 1926 distances, and it became
increasingly clear that they had further to be decreased to account
for Galactic absorption (cf.\ Table~\ref{tab:01}). But Hubble gave
his 1936 distances to E.~Holmberg as his preferred values still in
1950. 

   Yet Hubble's distance scale was challenged by \citet{Behr:51}. He
noticed the large luminosity scatter of Local Group galaxies and he
argued via the Malmquist effect that Hubble's {\em mean\/} luminosity
was too faint by $\sim\!1\fm5$ if applied to more distant,
magnitude-selected galaxies. (This is to our knowledge the first
mentioning of Malmquist statistics in extragalactic work; cf.\
Section~\ref{sec:07}). Citing \citet{Baade:44} he also corrected
Hubble's magnitudes by $0\fm35$ (at $18\fm3$). These were Behr's two
main reasons for deriving a value of $H_{0}=260$. He would have found
an even smaller value had he known of Stebbins, Whitford, \&
Johnson's (\citeyear{Stebbins:etal:50}) pioneering photoelectric
photometry which  proved Hubble's photometric scale error to be even
larger. 

   The revision received much more weight when \citet{Baade:52}
announced that work in M\,31 had shown, that either the
zero point of the Cepheids or of the RR\,Lyr stars must be in
error. Since Sandage's (published \citeyear{Sandage:53})
color-magnitude diagram of M\,3 had shown that the RR\,Lyr stars are
correct, the Cepheid luminosities had to be increased, as
\citet{Mineur:45} had already suggested. Baade concluded that
``previous estimates of extragalactic distances \dots were too small
by as much as a factor of 2'', which led him to $H_{0}\sim 250$.

   Accounting for the first four years of research with the $200''$
telescope, \citet{Sandage:54} summarized the evidence for $H_0$ and
concluded $125<H_{0}<276$.

   \citet{Humason:etal:56} estimated $H_{0}=180$ on two grounds: 
(1) They showed that what Hubble had considered as brightest stars of
NGC\,4321, a member of the Virgo cluster, were actually HII
regions. The brightest stars set in only $\sim\!2^{\rm m}$
fainter. 
(2) The absolute magnitude of M\,31, resulting from its apparent
Cepheid modulus of $(m-M)=24.25$ \citep{Baade:Swope:54}, could be used
by the authors to calibrate the upper-envelope line of their Hubble
diagram of field galaxies on the assumption that the luminosity of
M\,31 must be matched by at least some galaxies.

   The confusion between brightest stars and HII regions was
elaborated by \citet{Sandage:58}. The corresponding correction
together with the correction of Hubble's photometric scale led him to
conclude that the 1936 distance scale was too short by $4\fm6$ and
consequently that $H_{0}=75$. He noted that if the brightest stars had
$M_{\rm pg}=-9.5$ (which is now well demonstrated) $H_0$ would become
$55$. He also concluded from novae that Hubble's Local Group distances
were more nearly correct, i.e.\ too small by ``only'' $2\fm3$ on
average (cf.\ Table~\ref{tab:01}). Sandage's paper has become a
classic for not only having  given the first modern values of $H_0$,
but also because it contains the first physical explanation of the
instability strip of Cepheids.

   The situation in mid-1961 was summarized by \citet{Sandage:62} at
the influential 15th IAU Symposium (cf.\ Table~\ref{tab:01}). While he
cited values of $H_{0}\sim110$ by \citet{Sersic:60},
\citet{vandenBergh:60}, and \citet{Holmberg:58}, his own values ---
based, in addition to Cepheids and brightest stars, on the {\em
  size\/} of HII regions --- were 75$-$82. F.~Zwicky pleaded in the
discussion for $H_{0}=175$ from supernovae.  

\begin{table}[t]
\caption{Distances of nearby galaxies from 1936$-$2002.}
\label{tab:01}
\begin{center}
\begin{minipage}{\textwidth}

\bigskip
\begin{tabular}{lclccclclcl}
\hline
\noalign{\smallskip}
 & \multicolumn{3}{c}{Hubble} & & \multicolumn{2}{c}{Sandage}& & \\
 & 1936$^{*}$ & 1950 & $(m-M)^0$$^{**}$ & \hspace*{0.5cm} 
 & 1958$^{***}$ & 1962 & \hspace*{0.5cm} & 2002$^{\dagger}$ & \hspace*{0.5cm} 
 & Factor  \\
\noalign{\smallskip}
\hline
\noalign{\smallskip}
LMC       & 17.1 & 17.1 & 16.7 && 19.2 & 18.4  && 18.56 && 1.08 \\
SMC       & 17.3 & 17.3 & 17.0 && 19.2 & 18.78 && 19.00 && 1.11 \\
M\,31     & 21.6 & 22.4 & 21.8 && 24.6 & 24.16 && 24.47 && 1.14 \\
\noalign{\bigskip}
NGC\,2403 &      & 24.0 & 23.6 &&      & 27.6  && 27.65 && 1.02 \\
M\,101    &(25.5)& 24.0 & 23.6 &&      & 27.7  && 29.36 && 2.1  \\
Virgo     & 26.7 & 26.8 & 26.8 &&      & 30.8  && 31.6  && 1.4  \\
\noalign{\smallskip}
\hline
\noalign{\smallskip}
$H_0$     & 526  &      &      && 75 & 77$\pm$20 && 58$\pm$6 && \\ 
\noalign{\smallskip}
\hline
\noalign{\smallskip}
\end{tabular}

\medskip
$^{*}$ Hubble's 1926 distances are $0\fm5$ {\em larger} \hfill
$^{\dagger}$ as compiled by \citet{Tammann:etal:01b} \\
$^{**}$ corrected for galactic absorption \\
$^{***}$ from Novae
\end{minipage}
\end{center}
\end{table}

\section{A New Start on the Distance Scale 1960--1990}
\label{sec:05}
The still unchallenged (within the errors) Cepheid distance of M\,31
of $(m-M)^{0}=24.20\pm0.14$ \citep{Baade:Swope:63}, derived by H.H.
Swope after W. Baade's death from his $200''$-plates 
and from H.\,C. Arp's photoelectric sequence, may be considered as the
beginning of a new era. (For the history of the time cf.\ also
\citealt{Sandage:98,Sandage:99a}).

   By the same time the ``direct'' (i.e.\ non-spectroscopic) staff
members at the Mount Wilson and Palomar observatories (W. Baade,
E. Hubble, M. Humason, A. Sandage, and others) had accumulated many
$200''$-plates of a few galaxies outside the Local Group for work on
the Cepheids. In addition, A. Sandage had set up photoelectric
sequences around these galaxies, whose faintness and quality has
remained  unsurpassed until the advent of CCD detectors.

   Although the first Cepheid distance of NGC\,2403 to come out of
this program confirmed Sandage's \citeyear{Sandage:62} value
\citep{Tammann:Sandage:68}, using the then latest version of the
Cepheid period-luminosity (P-L) relation \citep{Sandage:Tammann:68},
it was criticized as being (much) too large
\citep[e.g.][]{Madore:76,deVaucouleurs:78,Hanes:82}. The modern value
is actually marginally larger \citep{Freedman:etal:01}.

   The second galaxy of the program, NGC\,5457 (M\,101), came as a
great surprise: its distance was found twice the value of Sandage's 
(\citeyear{Sandage:62}) estimate \citep{Sandage:Tammann:74a}, i.e.\
$(m-M)^{0}=29.3$. The distance of M\,101 and its companions was based on
brightest stars, HII region sizes, and van den Bergh's
(\citeyear{vandenBergh:60}) luminosity classes of spiral galaxies,
but also heavily on the {\em absence\/} of Cepheids down to the
detection limit. The faint Cepheids were eventually found with HST,
yielding  $(m-M)^{0}=29.34$ \citep{Kelson:etal:96}. In the mean time the
distance had been denounced as being too large
\citep[e.g.][]{deVaucouleurs:78,Humphreys:Strom:83}. 

   The new distance of M\,101 made clear that the brightest spirals of
luminosity class (LC)\,I  are brighter than anticipated and
that the luminosity of their brightest stars and the size of their
largest HII regions had to be increased. This led immediately to a
distance of the Virgo cluster of $(m-M)=31.45$
\citep{Sandage:Tammann:74b}, --- a value probably only slightly too
small \citep[cf.][]{Tammann:etal:01b}. The ensuing luminosity
calibration of LC\,I spirals could then be applied to a specially
selected, {\em distance-limited\/} sample of 36 such galaxies, bounded
by $8500\kms$. The conclusion was that $H_{0}=55\pm5$ ``everywhere''
\citep{Sandage:Tammann:75}. The largest contribution to the systematic
errors was attributed to the calibration through Cepheids (cf.\
Section~\ref{sec:09}). 

   The main antagonist of this solution became G. de
Vaucouleurs. Having started with $H_{0}=50$ from brightest globular
clusters \citep{deVaucouleurs:70}, he switched to $H_{0}\sim100\pm10$
\citep{deVaucouleurs:77,deVaucouleurs:Bollinger:79}. By assuming
rather short local distances and by turning a blind eye to all
selection effects, he managed to maintain this value --- eventually
with strong directional variations --- until his last paper on the
subject \citep{deVaucouleurs:Peters:85}.

   In the mean time new distance indicators were proposed which
correlate global galaxian observables with the total luminosity or
the diameter of a galaxy. An extrapolation of {\"O}pik's
(\citeyear{Oepik:21}) method was to use the 21cm HI-line width (a
measure of the rotational velocity) as a mass and hence luminosity
indicator, i.e.\ the so-called Tully-Fisher relation
\citep{Gouguenheim:69,Tully:Fisher:77,Sandage:Tammann:76}. It followed
the L-$\sigma$ \citep{Minkowski:62,Faber:Jackson:76}, D$_n$-$\sigma$
\citep{Dressler:etal:87}, or fundamental plane
\citep{Djorgovski:Davis:87} method and the surface brightness
fluctuation (SBF) method \citep{Tonry:Schneider:88}. Since these
relations have considerable intrinsic scatter their discussion is
deferred to Section~\ref{sec:07}.

   The turnover point of the bell-shaped luminosity function of
globular clusters has been used as a distance indicator
\citep*{vandenBergh:etal:85} . The successes and failures of the method
are discussed elsewhere \citep{Tammann:Sandage:99}.

   Following a proposal by \citet{Ford:Jenner:78} also brightest
planetary nebulae have been widely used as distance indicators. But
the method seems to depend on population size
\citep{Bottinelli:etal:91,Tammann:93}, chemical composition, and age
\citep{Mendez:etal:93}.

\section{A Complication: Peculiar Motions}
\label{sec:06}
Peculiar motions, addressed already by \citet{deSitter:30}, may
camouflage the true value of $H_0$, the negative velocity of M\,31
being a clear warning. Moreover \citet{Zwicky:33} had pointed out the
enormous virial motions in high-density clusters. But in the 1930's
the one-dimensional peculiar motions of field galaxies were estimated
to be $\la200\kms$ \citep{Hubble:Humason:31,Hubble:Humason:34}. 
Later work has shown that they are rather $\la50\kms$ on average at a
local scale of $5\;$Mpc 
\citep{Yahil:etal:77,Sandage:86,Ekholm:etal:01,Karachentsev:etal:02}
or even of $10\;$Mpc \citep{Tammann:Kraan-Korteweg:78}. Yet the size
of the peculiar motions increases with the scale length.

   The first physically motivated detection of a significant streaming
velocity, viz.\ towards the Virgo cluster center, is due to
Peebles (\citeyear{Peebles:76}; \citealt{Davis:Peebles:83}). 
Later work showed that the
Virgocentric flow amounts to $\sim\!220\pm50\kms$ at the position of the
Local Group \citep{Tammann:Sandage:85,Kraan-Korteweg:86}. Much larger
streaming velocities have since been proposed over still larger
scales, but it has never become clear, whether they are not due to Malmquist
bias of poorly defined galaxy samples (Section~\ref{sec:07}). 

   The total (three-dimensional) vector of $\sim\!630\kms$ of all
peculiar motions the Local Group partakes in, is reflected in the CMB
dipole \citep{Corey:Wilkinson:76, Smoot:etal:77}. A significant
fraction of the dipole is probably generated on scales within
$\la6000\kms$ \citep{Dale:etal:99,daCosta:etal:00}, although some
authors still favor higher values. As a consequence a reliable
large-scale value of $H_0$ should be determined at $>6000-10\,000\kms$
(Section~\ref{sec:08}).

   Notwithstanding the presence of large bulk motions the mean value
of $H_0$ turns out to be surprisingly scale-invariant. Within
$v\sim1200\kms$ (excluding the Virgo region) $H_0$ is only
insignificantly different from the large-scale value by $\Delta
H_{0}=1.8\pm2.7$ \citep[][Fig.~9]{Tammann:etal:01b}. Note that the
local and distant values of $H_0$ are about equally affected by any
errors of the underlying Cepheid calibration (cf.\ Section~\ref{sec:09}).

\section{Yet Another Problem: Selection Bias}
\label{sec:07}
The lesson of \citet{Malmquist:20} was only slowly incorporated by
extragalactic workers and is still not fully accounted for in some
modern work. The essence is that the mean luminosity of objects drawn
from an (apparent-) {\em magnitude-limited\/} sample increases with
distance, and that one overestimates $H_0$ as a consequence of this,
if one applies a local calibration to distant objects.

   The influence of the Malmquist bias was greatly exacerbated when
distance indicators came into use which use global galaxian
parameters as distance indicators, like the LC, TF, D$_{n}$-$\sigma$
(or fundamental-plane), or SBF methods. These methods have important
luminosity scatter in the order of $\sigma_{M}>0\fm3$, and since
the bias is strongly dependent on the size of the true intrinsic
scatter $\sigma_{M}$ (which is difficult to determine), they are
highly susceptible to Malmquist bias. Non-allowance for Malmquist bias
leads always, of course, to too high values of $H_0$.

   The literature abounds with $H_0$ determinations by means of the
above relations as applied to (complete or even incomplete) {\em
  magnitude-limited\/} samples. They typically lead to incorrect
values of $70 \la H_{0} \la 80$.

   The obvious way to overcome the Malmquist bias is to use {\em
distance-limited\/} samples. Examples are TF distances of a complete
galaxy sample bounded by $1000\kms$ (avoiding the wider Virgo region
as being too noisy for velocities to represent a good distance cutoff;
\citealt{Federspiel:99}; cf. \citealt{Tammann:etal:01b}, Fig.~5a) and
giving $H_{0}=59\pm3$, or a complete sample of Virgo cluster spirals
yielding a TF modulus of $(m-M)_{\rm Virgo}=31.58\pm0.24$
\citep{Federspiel:99}. 

   But complete distance- or volume-limited samples are restricted to
quite local regions. Several authors have therefore developed various
methods how to derive unbiased values of $H_0$ from (complete)
magnitude-limited samples of field galaxies
(\citealt{Teerikorpi:84,Teerikorpi:97,Bottinelli:etal:86};
\citealt*{Federspiel:etal:94}; 
for a tutorial see \citealt{Sandage:95}; for a parameter-free method
see \citealt{Hendry:02}). Also cluster samples are affected by
Teerikorpi Cluster Population Incompleteness Bias
\citep*{Teerikorpi:87,Sandage:etal:95}. The original hope that the
inverse TF relation was bias-free has not substantiated
\citep{Teerikorpi:etal:99}. 

   Bias-corrected samples of field galaxies give typical values of
$H_{0}\la60$, e.g. from LC
\citep{Sandage:96,Sandage:99b,Sandage:02,Paturel:etal:98}, from TF
\citep{Theureau:etal:97,Theureau:00,Ekholm:etal:99,Hendry:02}, from
galaxy diameters \citep{Goodwin:etal:97}, and from the
D$_{n}$-$\sigma$ method \citep{Federspiel:99}.

   All these solutions reach hardly beyond $5000\kms$, even in the
best cases. The determination of the large-scale value of $H_0$ is
therefore deferred to the next Section.

\section{The Golden Route to the Large-Scale Value of \boldmath{$H_0$}}
\label{sec:08}
Two developments have opened a simple and reliable way for the
determination of $H_0$ out to distances of $\sim\!30\,000\kms$.

\noindent
1) Following an early suggestion by \citet{Kowal:68} it has become
increasingly clear that SNe\,Ia are the most powerful standard
candles to date. The scatter of their Hubble diagram decreased
steadily as it became possible: a) to remove non-type Ia SNe from the
sample, b) to obtain good photometry of their magnitudes at maximum,
many of which are due to a specific program at Cerro Tololo, c) to
allow for the dependence of $m_{\max}$ on the decline rate of the
light curve, --- a correction which turned out to be smaller than
originally proposed by \citet{Phillips:93}, and d) to correct for a
dependence of $m_{\max}$ on the intrinsic color of the SN\,Ia
\citep{Tammann:Sandage:95,Tripp:98}. \citet{Parodi:etal:00} have
compiled a complete fiducial sample of 35 blue SNe\,Ia, which fulfill
the conditions $(B\!-\!V)^0 \le 0.10$ and
$1200<v<30\,000\kms$ (the lower limit is imposed to reduce the effect
of peculiar velocities, the upper limit is to minimize the effect of
$\Omega_{\rm M}$ and $\Omega_{\Lambda}$). This sample allows to
determine good dependencies of $m_{\max}$ on decline rate and
intrinsic color. The resulting $m_{\max}^{\rm corr}$ in $B$, $V$, and
$I$ define very tight Hubble diagrams
\citep[cf.][]{Parodi:etal:00}. As an example only the Hubble diagram
in $V$ is shown here (Fig.~\ref{fig:hub:sneia}). 
The solution for the Hubble line, shown in the Figure, excludes the
nine SNe\,Ia which have a Galactic absorption of $A_{V}>0\fm2$
according to \citet{Schlegel:etal:98}.  They appear to be somewhat
bright and their absorption may be overestimated.
\def\floatwidth{0.7\textwidth} 
\begin{figure}[t] 
\centerline{\includegraphics[width=\floatwidth]{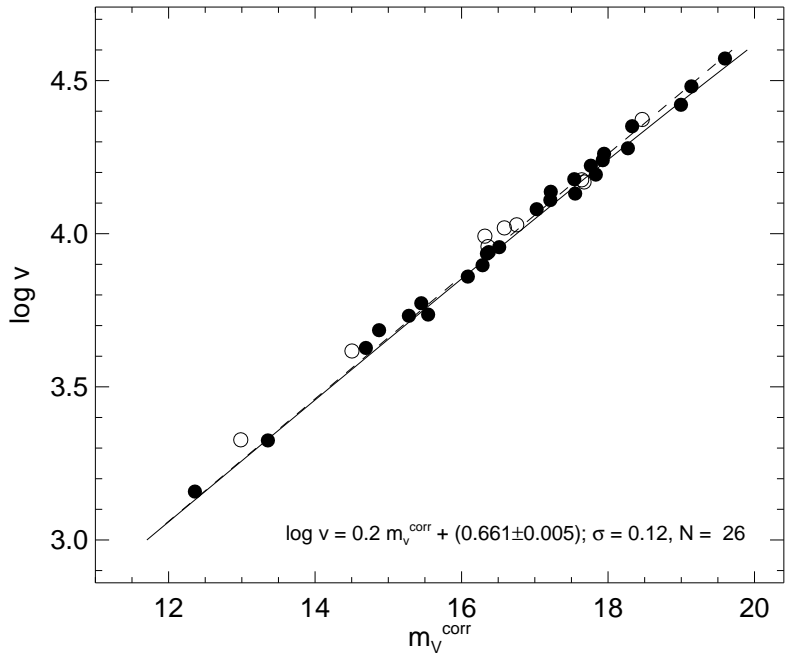}} 
\caption{The Hubble diagram in $V$ for the 35 
  SNe\,Ia of the fiducial sample with magnitudes $m^{\rm corr}_V$, 
  i.e. corrected for decline rate $\Delta m_{15}$ and color $(B-V)$. 
   The solid line is a fit to the 26 SNe\,Ia with 
  $A_V\le0\fm2$ (closed circles) 
  assuming a flat universe with $\Omega_{\rm M}=0.3$ and 
  $\Omega_{\Lambda}=0.7$; the dashed line is a linear fit with a
  forced   slope of 0.2 (corresponding approximately to $\Omega_{\rm
  M}=1.0$ and $\Omega_{\Lambda}=0.0$).} 
\label{fig:hub:sneia}
\end{figure}

   The very small scatter of $\sigma_{m}=0\fm12$ in
Fig.~\ref{fig:hub:sneia} could be entirely due to observational errors
(in magnitude and color), to which peculiar velocities may contribute
to some extent. The luminosity variation of homogenized, blue SNe\,Ia
is therefore below the detection limit. Their unique r{\^o}le as
standard candles and their insensitivity against Malmquist bias are
unquestionable. 

   The conclusion is that SNe\,Ia open the golden route to the
large-scale value of $H_0$ if ``only'' their absolute magnitude
$M^{\rm corr}$ can be determined. 

\noindent
2) The advent of HST made it possible --- as a pilot program had shown
\citep{Sandage:Tammann:82} --- that the required luminosity
calibration of SNe\,Ia could be achieved. A special HST program was
mounted with the aim of determining Cepheid distances of {\em
  nearby\/} galaxies which have produced a well observed SN\,Ia in
the past. This has now led to the distances of seven standard SNe\,Ia,
which were  augmented by two additional objects from external
sources. The data are compiled in \citet{Saha:etal:01}. The mean
absolute magnitude of the nine SNe\,Ia, after correction for decline
rate and color, becomes $M_{V}^{\rm corr}=-19.53\pm0.06$. 

   If this value is combined with the equation of the corresponding
Hubble line, as shown at the bottom of Fig.~\ref{fig:hub:sneia},
one obtains 
\begin{equation}\label{eq:h0}
   H_{0} = 56.9\pm2.3.
\end{equation}
Very similar solutions are obtained from the $B$ and $I$ magnitudes
\citep[cf.][]{Tammann:etal:01b}. The value of $H_0$ could be increased
by 1.6 units if the LMC distance modulus, i.e.\ the zero point of the
Cepheid distance scale, were adopted to be 18.50, instead of 18.56
used here. It could also be increased by 1.3 units if all 35 SNe\,Ia
were included in the solution. $H_0$ decreases by 0.8 units for an
$\Omega_{\rm M}=1$, $\Omega_{\Lambda}=0$ Universe. (Note the
  non-vanishing effect which $\Lambda$ has already at
  $v=30\,000\kms$).

   An important feature of the present solution is that the
calibrating SNe\,Ia and the distant sample of SNe\,Ia have the same
absorption-corrected mean colors. Solutions which do not fulfill this
condition betray an inconsistent treatment of the two sets.

   Equation~(\ref{eq:h0}) reflects only statistical errors. The
systematic errors are discussed by \citet{Saha:etal:01} and
\citet{Tammann:etal:01b}. The conclusion is that the systematic error
of $H_0$ is $\la10\%$, the slope of the magnitude dependence on
decline rate and color contributing somewhat, but the main source of
error being the Cepheid distances discussed in the next Section. 

   A note is here in place on the equally tight Hubble diagram from
relative TF distances of clusters (\citealt{Giovanelli:etal:97};
\citealt{Dale:etal:99}; cf. \citealt{Tammann:etal:01a}, Fig.~2),
where the cluster distances are based on the mean of about 15
subjectively selected spirals per cluster. The formidable problem is,
however, how to calibrate the diagram. One possibility is to {\em
  adopt\/} the distance of one or more clusters. 
\citet{Sakai:etal:00} have applied instead the local calibration of
the TF relation (derived from known Cepheid distances) directly to the
individual cluster galaxies, and have obtained a value of $H_{0}=71$. 
This procedure is of course invalid. It is well known that if one
applies the TF calibration to only some bright members of the Virgo
cluster or any other cluster, the mean distance 
becomes $\sim\!0\fm4$ smaller than from the complete cluster spiral
population \citep[Fig.~6]{Kraan-Korteweg:etal:88}.
This is a typical example of the Teerikorpi Cluster Bias effect (cf.\
Section~\ref{sec:07}). It is therefore obvious that the value of $H_0$
by \citeauthor{Sakai:etal:00} must be too high by roughly $20\%$.

   The same objection holds against the calibration of the Hubble
diagram of 11 clusters, whose relative distances are based on the mean
D$_n-\sigma$ distances of a few cluster E/S0 members
\citep{Kelson:etal:00}. The additional problem is here that even the
local calibration of the D$_n-\sigma$ relation is shaky because there
is no early-type galaxy with a primary distance determination.

   Purely physical methods leading directly to $H_0$, without using
any astronomical distance for calibration, are steadily gaining
weight. A combined solution of six gravitationally lensed double or
multiple quasars yields for a $\Lambda$CDM model a well constrained
value of $50<H_{0}<60$ \citep{Saha:02}. The physics of the light curve
plateau of SNe of type IIP suggests values of $54\la H_{0}\la65$
\citep{Hamuy:01,Nadyozhin:02}. \citet{Lasenby:02}, summarizing the
data from the Sunyaev-Zeldovich effect, concludes that $H_0$ is
``sixtyish''. CMB fluctuations are compatible with the same estimate
\citep{Netterfield:etal:02,Pryke:02}, but instead of determing $H_0$
from the CMB it should rather be used as a prior to confine other
parameters of the early Universe \citep{Durrer:02}. This is in fact
the main motivation at present to determine as precise a
CMB-independent value of $H_0$ as possible.

\section{Back to the Beginnings}
\label{sec:09}
It was suggested \citep{Freedman:etal:01} that the absorption of LMC
Cepheids, widely used as zero point for all Cepheid distances, had
been overestimated so far and that Cepheids were intrinsically dimmer
than anticipated; this would result in an {\em increase\/} of $H_0$ by
$\sim\!6\%$. This suggestion is not correct for long-period Cepheids
\citep{Tammann:etal:01b}, which are decisive for the luminosity
calibration of SNe\,Ia, but the suggestion invited a closer look into
the P-L and P-C relations of Cepheids in different galaxies.

   The data for nearby Cepheids have dramatically increased in recent
years. \citet{Berdnikov:etal:00} have provided $B,V,I$ photometry for
hundreds of Galactic Cepheids for which \citet{Fernie:etal:95} have
determined reddening values. \citet{Feast:99} has compiled the
distances of 28 Cepheids in Galactic clusters, and
\citet{Gieren:etal:98} have determined distances of 34 Galactic
Cepheids by means of the Baade-Becker-Wesselink
method. \citet{Udalski:etal:99a,Udalski:etal:99b} have published
periods and magnitudes in the standard system, obtained in the course
of the OGLE program, for many hundreds of fundamental-mode Cepheids in
LMC and SMC as well as their consistent $E(B\!-\!V)$ values.
\def\floatwidth{0.64\textwidth}
\begin{figure}[tb]
\centerline{\includegraphics[width=\floatwidth]{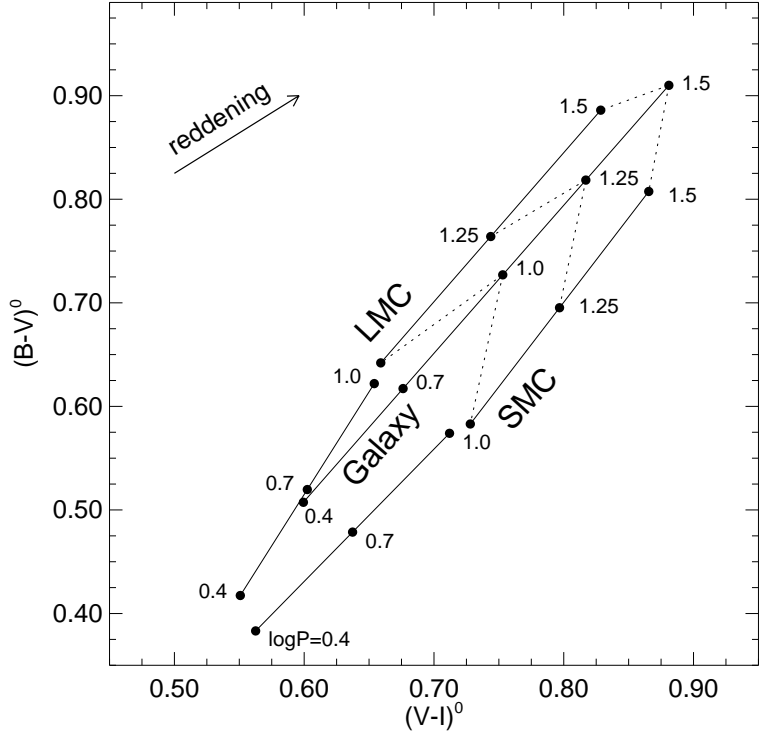}}
\caption{The mean position in the two-color diagram [$(B\!-\!V)^0$
  versus $(V\!-\!I)^0$] of Galactic, LMC, and SMC Cepheids. The loci
  of different values of $\log P$ are indicated.} 
\label{fig:2color}
\end{figure}

   The analysis of this wealth of data is still in progress, but it is
already clear that the P-L and P-C relations are different in
different galaxies. In LMC neither the P-L nor the P-C relation can be
fitted by a single slope \citep{Tammann:etal:01b}, whereas the steep
Galactic P-L relation shows no deviation from linearity. 
Cepheids in LMC are up to $0\fm5$
brighter at $\log P = 0.4$ than their Galactic counterparts (somewhat
dependent on the adopted distances), but the difference diminishes
towards longer periods. It is a lucky coincidence that the Galactic
and LMC P-L relations cross at $\log P \approx 1.5$ and that this is
about the median period of the SNe\,Ia-calibrating Cepheids. But the
choice of any specific P-L relation may influence the resulting
absorption-corrected distance of external galaxies by up to $10\%$. 

   A distance-independent comparison of the Cepheids in the Galaxy,
LMC, and SMC is afforded by their mean position in the color-color
diagram (Fig.~\ref{fig:2color}). It cannot entirely be excluded that the
(large) Galactic $E(B\!-\!V)$ values are underestimated and that the
Galactic Cepheids should be slid closer to the LMC Cepheids along the
reddening line, but the fact that Cepheids in SMC at a given period
are bluer in $(B\!-\!V)^0$ and redder in $(V\!-\!I)^0$ than those in
LMC cannot be the result of incorrect reddenings. If the P-C relations
are not universal then it follows by necessity that also the P-L
relations must be different, at least at some wavelengths.

   It comes as a surprise that the largest single source of systematic
errors, which affect the luminosity calibration of SNe\,Ia and hence
the large-scale value of $H_0$, is due to the intrinsic difference of
Cepheids in different galaxies. {\em Of course, all other distance
determinations which are calibrated through Cepheids face the same
problem}. At present it is not known which parameter decides on the
specific form of the P-L relation, but metallicity is a prime
suspect. If this is so, and considering that the mean metallicity of
the SNe\,Ia-calibrating galaxies is roughly $\mbox{[Fe/H]}=0$, one
should prefer the Galactic over the LMC P-L relation. In that case the
galaxy distances, based on long-period Cepheids, would tend to
increase.

   It is a long-range program of the future to trace accurate P-C and
calibrated P-L relations in a variety of nearby galaxies, to understand
the physics of their being different, and then to decide about their
applicability to more distant galaxies. It may take the astrometric
distances from GAIA to calibrate the various P-L relations for further
use as distance indicators. At present it is prudent to attribute a
systematic error of at least $5\%$ to the value of $H_0$, accounting
for only the intrinsic difference of Cepheids.

\section*{Acknowledgments}
The authors thank Dr.~A.~Sandage for the permission to use his
manuscript on the history of the Mount Wilson Observatory. They thank
the Swiss National Science Foundation for financial support.



\begin{thebibliography}{}
%
\bibitem[Baade(1944)]{Baade:44}
   Baade, W. 1944,
   {\em ApJ} {\bf 100}, 137.
%
\bibitem[Baade(1952)]{Baade:52}
   Baade, W. 1952,
   {\em I.A.U. Trans.} VIII (Cambridge: Cambridge Univ. Press), p. 397.
%
\bibitem[Baade(1963)]{Baade:63}
   Baade, W. 1963,
   {\em Evolution of Stars and Galaxies}, ed. C. Payne-Gaposchkin,
   (Cambridge, Mass.: Harvard Univ. Press), p.~28. 
%
\bibitem[Baade \& Swope(1954)]{Baade:Swope:54}
   Baade, W., \& Swope, H.H. 1954,
   {\em Yearbook Carnegie Inst.} {\bf 53}, 20.
%
\bibitem[Baade \& Swope(1963)]{Baade:Swope:63}
   Baade, W., \& Swope, H.H. 1963,
   {\em AJ} {\bf 68}, 435.
%
\bibitem[Behr(1951)]{Behr:51}
   Behr, A. 1951,
   {\em A.N.} {\bf 279}, 97.
%
\bibitem[Berdnikov et~al.(2000)Berdnikov, Dambis, \&
   Voziakova]{Berdnikov:etal:00} 
   Berdnikov, L.N., Dambis, A.K., \& Voziakova, O.V., 2000,
   {\em A\&AS} {\bf 143}, 211.
%
\bibitem[Bottinelli et~al.(1986)]{Bottinelli:etal:86}
   Bottinelli, L., Gouguenheim, L., Paturel, G., \& Teerikorpi,
   P. 1986,
   {\em A\&A} {\bf 156}, 157.
%
\bibitem[Bottinelli et~al.(1991)]{Bottinelli:etal:91}
   Bottinelli, L., Gouguenheim, L., Paturel, G., \& Teerikorpi,
   P. 1991,
   {\em A\&A} {\bf 252}, 550.
%
\bibitem[Corey \& Wilkinson(1976)]{Corey:Wilkinson:76}
   Corey, B.E., \& Wilkinson, D.T. 1976,
   {\em BAAS} {\bf 8}, 351.
%
\bibitem[da Costa et~al.(2000)]{daCosta:etal:00}
   da Costa, L.N., et~al. 2000,
   {\em ApJ} {\bf 537}, L81.
%
\bibitem[Dale et~al.(1999)]{Dale:etal:99}
   Dale, D.A., Giovanelli, R., Haynes, M.P., Campusano, L.E., \&
   Hardy, E. 1999,
   {\em AJ} {\bf 118}, 1489.
%
\bibitem[Davis \& Peebles(1983)]{Davis:Peebles:83}
   Davis, M., \& Peebles, P.J.E. 1983,
   {\em ARA\&A} {\bf 21}, 109.
%
\bibitem[de Sitter(1916)]{deSitter:16}
   de Sitter, W. 1916,
   {\em MNRAS} {\bf 76}, 699.
%
\bibitem[de Sitter(1917)]{deSitter:17}
   de Sitter, W. 1917,
   {\em MNRAS} {\bf 78}, 3.
%
\bibitem[de Sitter(1930)]{deSitter:30}
   de Sitter, W. 1930,
   {\em BAN} {\bf 5}, 157.
%
\bibitem[de Vaucouleurs(1970)]{deVaucouleurs:70}
   de Vaucouleurs, G. 1970,
   {\em ApJ} {\bf 159}, 435.
%
\bibitem[de Vaucouleurs(1977)]{deVaucouleurs:77}
   de Vaucouleurs, G. 1977,
   in: {\em D{\'e}calage vers le rouge et expansion de l'univers},
   eds. C. Balkowski \&  B.E. Westerlund, IAU Coll. {\bf 37}, (Paris:
   Ed. CNRS), p.~301. 
%
\bibitem[de Vaucouleurs(1978)]{deVaucouleurs:78}
   de Vaucouleurs, G. 1978,
   {\em ApJ} {\bf 224}, 710.
%
\bibitem[de Vaucouleurs \& Bollinger(1979)]{deVaucouleurs:Bollinger:79}
   de Vaucouleurs, G., \& Bollinger, G. 1979,
   {\em ApJ} {\bf 233}, 433.
%
\bibitem[de Vaucouleurs \& Peters(1985)]{deVaucouleurs:Peters:85}
   de Vaucouleurs, G., \& Peters, W. 1985,
   {\em ApJ} {\bf 297}, 27.
%
\bibitem[Djorgovski \& Davis(1987)]{Djorgovski:Davis:87}
   Djorgovski, S., \& Davis, M. 1987,
   {\em ApJ} {\bf 313}, 59.
%
\bibitem[Dressler et~al.(1987)]{Dressler:etal:87}
   Dressler, A., Faber, S.M., Burstein, D., Davies, R.L., Lynden-Bell,
   D., Terlevich, R.J., \& Wegner, G. 1987,
   {\em ApJ} {\bf 313}, L37.
%
\bibitem[Durrer(2002)]{Durrer:02}
   Durrer, R. 2002,
   in: {\em Matter in the Universe}, eds. Ph. Jetzer, K. Pretzl, \&
   R. von Steiger, (Dordrecht: Kluwer), p.~3.
%
\bibitem[Ekholm et~al.(1999)]{Ekholm:etal:99}
   Ekholm, T., Teerikorpi, P., Theureau, G., Hanski, M., Paturel, G.,
   Bottinelli, L., \& Gouguenheim, L. 1999,
   {\em A\&A} {\bf 347}, 99.
%
\bibitem[Ekholm et~al.(2001)]{Ekholm:etal:01}
   Ekholm, T., Baryshev, Y., Teerikorpi, P., Hanski, M., \&  Paturel,
   G. 2001,
   A\&A {\bf 368}, 17.
%
\bibitem[Faber \& Jackson(1976)]{Faber:Jackson:76}
   Faber, S.M., \& Jackson, R.E. 1976,
   {\em ApJ} {\bf 204}, 668.
%
\bibitem[Feast(1999)]{Feast:99}
   Feast, M.W. 1999,
   {\em PASP} {\bf 111}, 775.
%
\bibitem[Federspiel(1999)]{Federspiel:99}
   Federspiel, M. 1999,
   Ph.D. Thesis, Univ. Basel.
%
\bibitem[Federspiel et~al.(1994)Federspiel, Sandage, \&
Tammann]{Federspiel:etal:94} 
   Federspiel, M., Sandage, A., \& Tammann, G.A. 1994,
   {\em ApJ} {\bf 430}, 29.
%
\bibitem[Fernie(1970)]{Fernie:70}
   Fernie, J.D. 1995,
   {\em PASP} {\bf 82}, 1189.
%
\bibitem[Fernie et~al.(1995)]{Fernie:etal:95}
   Fernie, J.D., Beattie, B., Evans, N.R., \& Seager, S. 1995,
   {\em IBVS} {\bf 4148}. 
%
\bibitem[Ford \& Jenner(1978)]{Ford:Jenner:78}
   Ford, H.C., \& Jenner, D.C. 1978,
   {\em BAAS} {\bf 10}, 665.
%
\bibitem[Freedman et~al.(2001)]{Freedman:etal:01}
   Freedman, W.L., et~al. 2001,
   {\em ApJ} {\bf 553}, 47.
%
\bibitem[Gieren et~al.(1998)Gieren, Fouqu{\'e}, \& Gomez]{Gieren:etal:98}
   Gieren, W.P., Fouqu{\'e}, P., \& Gomez, M. 1998,
   {\em ApJ} {\bf 496}, 17.
%
\bibitem[Giovanelli et~al.(1997)]{Giovanelli:etal:97}
   Giovanelli, R., et~al. 1997,
   {\em AJ} {\bf 113}, 22.
%
\bibitem[Goodwin et~al.(1997)]{Goodwin:etal:97}
   Goodwin, S.P., Gribbin, J., \& Hendry, M.A. 1997,
   {\em AJ} {\bf 114}, 2212.
%
\bibitem[Hamuy(2001)]{Hamuy:01}
   Hamuy, M.A. 2001,
   Ph.D. Thesis, Univ. of Arizona.
%
\bibitem[Gouguenheim(1969)]{Gouguenheim:69}
   Gouguenheim, L. 1969,
   {\em A\&A} {\bf 3}, 281 (eq. 11c).
%
\bibitem[Hanes(1982)]{Hanes:82}
   Hanes, D.A. 1982,
   {\em MNRAS} {\bf 201}, 145.
%
\bibitem[Hendry(2002)]{Hendry:02}
   Hendry, M. 2002,
   in: {\em A New Era in Cosmology}, eds. T. Shanks \& N. Metcalfe,
   ASP Conf. Ser., in press.
%
\bibitem[Hertzsprung(1914)]{Hertzsprung:14}
   Hertzsprung, E. 1913,
   {\em A.N.} {\bf 196}, 201.
%
\bibitem[Holmberg(1958)]{Holmberg:58}
   Holmberg, E. 1958,
   {\em Medd. Lunds Astr. Obs.} {\bf 2}, No.\,136.
%
\bibitem[Hubble(1925)]{Hubble:25}
   Hubble, E. 1925,
   {\em ApJ} {\bf 62}, 409.
%
\bibitem[Hubble(1926)]{Hubble:26}
   Hubble, E. 1926,
   {\em ApJ} {\bf 63}, 236.
%
\bibitem[Hubble(1929a)]{Hubble:29a}
   Hubble, E. 1929a,
   {\em ApJ} {\bf 69}, 103.
%
\bibitem[Hubble(1929b)]{Hubble:29b}
   Hubble, E. 1929b,
   {\em Proc. Nat. Acad. Sci.} {\bf 15}, 168.
%
\bibitem[Hubble(1936a)]{Hubble:36a}
   Hubble, E. 1936a,
   {\em The Realm of Nebulae} (New Haven: Yale Univ. Press).
%
\bibitem[Hubble(1936b)]{Hubble:36b}
   Hubble, E. 1936b,
   {\em ApJ} {\bf 84}, 517.
%
\bibitem[Hubble(1953)]{Hubble:53}
   Hubble, E. 1953,
   {\em MNRAS} {\bf 113}, 658.
%
\bibitem[Hubble \& Humason(1931)]{Hubble:Humason:31}
   Hubble, E., \& Humason, M.L. 1931,
   {\em ApJ} {\bf 74}, 43.
%
\bibitem[Hubble \& Humason(1934)]{Hubble:Humason:34}
   Hubble, E., \& Humason, M.L. 1934,
   {\em Proc. Nat. Acad. Sci. USA} {\bf 20}, 264.
%
\bibitem[Humason(1936)]{Humason:36}
   Humason, M.L. 1936,
   {\em ApJ} {\bf 83}, 10.
%
\bibitem[HMS(1956)Humason, Mayall, \& Sandage]{Humason:etal:56}
   Humason, M.L., Mayall, N.U., \& Sandage, A.R. 1956,
   {\em AJ} {\bf 61}, 97 (HMS).
%
\bibitem[Humphreys \& Strom(1983)]{Humphreys:Strom:83}
   Humphreys, R.M., \& Strom, S.E. 1983,
   {\em ApJ} {\bf 264}, 458.
%
\bibitem[Karachentsev et~al.(2002)]{Karachentsev:etal:02}
   Karachentsev, et~al. 2002,
   {\em A\&A} {\bf 389}, 812.
%
\bibitem[Kelson et~al.(1996)]{Kelson:etal:96}
   Kelson, et~al. 1996,
   {\em ApJ} {\bf 463}, 26.
%
\bibitem[Kelson et~al.(2000)]{Kelson:etal:00}
   Kelson, et~al. 2000,
   {\em ApJ} {\bf 529}, 768.
%
\bibitem[Kowal(1968)]{Kowal:68}
   Kowal, C.T. 1968,
   {\em AJ} {\bf 73}, 1021.
%
\bibitem[Kraan-Korteweg(1986)]{Kraan-Korteweg:86}
   Kraan-Korteweg, R.C. 1986,
   {\em A\&AS} {\bf 66}, 255.
%
\bibitem[Kraan-Korteweg, Cameron, \& Tammann(1988)]{Kraan-Korteweg:etal:88}
   Kraan-Korteweg, R.C., Cameron, L.M., \& Tammann, G.A. 1988,
   {\em ApJ} {\bf 331}, 620.
%
\bibitem[Lasenby(2002)]{Lasenby:02}
   Lasenby, A. 2002,
   Talk presented at a Coll., Basel 26.06.2002.
%
\bibitem[Lema{\^i}tre(1927)]{Lemaitre:27}
   Lema{\^i}tre, G. 1927,
   {\em Ann. Soc. Sci. Bruxelles} {\bf 47}, 49. (English translation
   in {\em MNRAS} {\bf 91}, 483 [1931]).
%
\bibitem[Lundmark(1920)]{Lundmark:20}
   Lundmark, K. 1920,
   {\em The relation of the globular clusters and spiral nebulae to
   the stellar system},
   (Stockholm: Handlingar).
%
\bibitem[Lundmark(1925)]{Lundmark:25}
   Lundmark, K. 1925,
   {\em MNRAS} {\bf 85}, 865.
%
\bibitem[Madore(1976)]{Madore:76}
   Madore, B.F. 1976,
   {\em MNRAS} {\bf 177}, 157.
%
\bibitem[Malmquist(1920)]{Malmquist:20}
   Malmquist, G. 1920,
   {\em Medd. Lunds Astr. Obs.} {\bf 2}, No.\,22.
%
\bibitem[M{\'e}ndez et~al.(1993)]{Mendez:etal:93}
   M{\'e}ndez, R.H., Kudritzki, R.P.,  Ciardullo, R., \&  Jacoby, G.H.  1993,
   {\em A\&A} {\bf 275}, 534.
%
\bibitem[Minkowski(1962)]{Minkowski:62}
   Minkowski 1962,
   in: {\em Problems of Extra-Galactic Research}, ed. G.C. McVittie,
   IAU Symp. {\bf 15}, (New York: Macmillan), p.~112.
%
\bibitem[Mineur(1945)]{Mineur:45}
   Mineur, H. 1945,
   {\em C.\,R. Acad. Sci. Paris} {\bf 220}, 445.
%
\bibitem[Nadyozhin(2002)]{Nadyozhin:02}
   Nadyozhin, D.K., 2002,
   in: {\em From Twilight to Highlight, the Physics of Supernovae},
   ESO/MPA/MPE Workshop, in press.
%
\bibitem[Netterfield et~al.(2002)]{Netterfield:etal:02}
   Netterfield, C.B., et~al. 2002,
   {\em ApJ} {\bf 571}, 604.
%
\bibitem[{\"O}pik(1921)]{Oepik:21}
   {\"O}pik, E. 1921,
   {\em Weltkunde} (Russian) {\bf 10}, 12.
%
\bibitem[{\"O}pik(1922)]{Oepik:22}
   {\"O}pik, E. 1922,
   {\em ApJ} {\bf 55}, 406.
%
\bibitem[Parodi et~al.(2000)]{Parodi:etal:00}
   Parodi, B.R., Saha, A., Sandage, A., \& Tammann, G.A. 2000,
   {\em ApJ} {\bf 540}, 634.
%
\bibitem[Paturel et~al.(1998)]{Paturel:etal:98}
   Paturel, G., et~al. 1998,
   {\em A\&A} {\bf 339}, 671.
%
\bibitem[Peebles(1976)]{Peebles:76}
   Peebles, P.J.E. 1976,
   {\em ApJ} {\bf 205}, 318.
%
\bibitem[Phillips(1993)]{Phillips:93}
   Phillips, M.M. 1993,
   {\em ApJ} {\bf 413}, L105.
%
\bibitem[Pryke(2002)]{Pryke:02}
   Pryke, C. 2002,
   in: {\em A New Era in Cosmology}, eds. T. Shanks \& N. Metcalfe,
   ASP Conf. Ser., in press.
%
\bibitem[Ritchey(1917)]{Ritchey:17}
   Ritchey, G.W. 1917,
   {\em PASP} {\bf 29}, 207.
%
\bibitem[Robertson(1928)]{Robertson:28}
   Robertson, H.P. 1928,
   {\em Phil. Mag.} {\bf 5}, 835.
%
\bibitem[Saha et~al.(2001)]{Saha:etal:01}
   Saha, A., Sandage, A., Tammann, G.A., Dolphin, A.E., Christensen,
   J., Panagia, N., \& Macchetto, E.D. 2001,
   {\em ApJ} {\bf 562}, 314.
%
\bibitem[Saha(2002)]{Saha:02}
   Saha, P. 2002,
   Talk presented at a Coll., Basel 26.06.2002.
%
\bibitem[Sakai et~al.(2000)]{Sakai:etal:00}
   Sakai, et~al. 2000,
   {\em ApJ} {\bf 529}, 698.
%
\bibitem[Sandage(1953)]{Sandage:53}
   Sandage, A. 1953,
   {\em AJ} {\bf 58}, 61.
%
\bibitem[Sandage(1954)]{Sandage:54}
   Sandage, A. 1954,
   {\em AJ} {\bf 59}, 180.
%
\bibitem[Sandage(1958)]{Sandage:58}
   Sandage, A. 1958,
   {\em ApJ} {\bf 127}, 513.
%
\bibitem[Sandage(1962)]{Sandage:62}
   Sandage, A. 1962,
   in: {\em Problems of Extra-Galactic Research}, ed. G.C. McVittie,
   IAU Symp. {\bf 15}, (New York: Macmillan), p.~359.
%
\bibitem[Sandage(1972)]{Sandage:72}
   Sandage, A. 1972,
   {\em ApJ} {\bf 178}, 1.
%
\bibitem[Sandage(1986)]{Sandage:86}
   Sandage, A. 1986,
   {\em ApJ} {\bf 307}, 1.
%
\bibitem[Sandage(1995)]{Sandage:95}
   Sandage, A. 1995,
   in: {\em The Deep Universe}, 
   Saas-Fee Advanced Course {\bf 23}, 
   eds. B. Binggeli \& R. Buser, 
   (Berlin: Springer), p.~1.
%
\bibitem[Sandage(1996)]{Sandage:96}
   Sandage, A. 1996,
   %
   {\em AJ} {\bf 111}, 1 and 18.
%
\bibitem[Sandage(1998)]{Sandage:98}
   Sandage, A. 1998,
   in: {\em Supernovae and Cosmology}, 
   eds. L. Labhardt, B. Binggeli, \& R. Buser, 
   (Basel: Astron. Inst.), p.~201.
%
\bibitem[Sandage(1999a)]{Sandage:99a}
   Sandage, A. 1999a,
   {\em ARA\&A} {\bf 37}, 445.
%
\bibitem[Sandage(1999b)]{Sandage:99b}
   Sandage, A. 1999b,
   {\em ApJ} {\bf 527}, 479.
%
\bibitem[Sandage(2002)]{Sandage:02}
   Sandage, A. 2002,
   {\em AJ} {\bf 123}, 1179.
%
\bibitem[Sandage \& Hardy(1973)]{Sandage:Hardy:73}
   Sandage, A., \& Hardy, E. 1973,
   {\em ApJ} {\bf 183}, 743.
%
\bibitem[Sandage et~al.(1976)Sandage, Kristian, \& Westphal]{Sandage:etal:76}
   Sandage, A., Kristian, J., \& Westphal, J.A. 1976,
   {\em ApJ} {\bf 205}, 688.
%
\bibitem[Sandage \& Tammann(1968)]{Sandage:Tammann:68}
   Sandage, A., \& Tammann, G.A. 1968,
   {\em ApJ} {\bf 151}, 531.
%
\bibitem[Sandage \& Tammann(1974a)]{Sandage:Tammann:74a}
   Sandage, A., \& Tammann, G.A. 1974a,
   {\em ApJ} {\bf 194}, 223.
%
\bibitem[Sandage \& Tammann(1974b)]{Sandage:Tammann:74b}
   Sandage, A., \& Tammann, G.A. 1974b,
   {\em ApJ} {\bf 194}, 559.
%
\bibitem[Sandage \& Tammann(1975)]{Sandage:Tammann:75}
   Sandage, A., \& Tammann, G.A. 1975,
   {\em ApJ} {\bf 197}, 265.
%
\bibitem[Sandage \& Tammann(1976)]{Sandage:Tammann:76}
   Sandage, A., \& Tammann, G.A. 1976,
   {\em ApJ} {\bf 210}, 7. 
%
\bibitem[Sandage \& Tammann(1982)]{Sandage:Tammann:82}
   Sandage, A., \& Tammann, G.A. 1982,
   {\em ApJ} {\bf 256}, 339.
%
\bibitem[Sandage et~al.(1995)Sandage, Tammann, \&
Federspiel]{Sandage:etal:95} 
   Sandage, A., Tammann, G.A., \& Federspiel, M. 1995,
   {\em ApJ} {\bf 452}, 1.
%
\bibitem[Schlegel et~al.(1998)Schlegel, Finkbeiner, \& Davis]{Schlegel:etal:98}
   Schlegel, D., Finkbeiner, D., \& Davis, M. 1998,
   {\em ApJ} {\bf 500}, 525.
%
\bibitem[Shapley(1915)]{Shapley:15}
   Shapley, H. 1915,
   {\em Mt. Wilson Contrib.} {\bf 116}, 82.
%
\bibitem[Shapley(1918)]{Shapley:18}
   Shapley, H. 1918,
   {\em ApJ} {\bf 48}, 154.
%
\bibitem[S{\'e}rsic(1960)]{Sersic:60}
   S{\'e}rsic, J.L. 1960,
   {\em Zs. f. Ap.} {\bf 50}, 168.
%
\bibitem[Slipher(1914)]{Slipher:14}
   Slipher, V.M. 1914,
   {\em Bulletin Lowell Obs.} {\bf 62}, 66.
%
\bibitem[Smoot, Gorenstein, \& Muller(1977)]{Smoot:etal:77}
   Smoot, G.F., Gorenstein, M.V., \& Muller, R.A. 1977, 
   {\em Phys. Rev. Lett.} {\bf 39}, 898.

%
\bibitem[Stebbins et~al.(1950)Stebbins, Whitford \& Johnson]{Stebbins:etal:50}
   Stebbins, J., Whitford, A.E., \& Johnson, H.L. 1950,
   {\em ApJ} {\bf 112}, 469.
%
\bibitem[Stromberg(1925)]{Stromberg:25}
   Stromberg, G. 1925,
   {\em ApJ} {\bf 61}, 353.
%
\bibitem[Tammann(1993)]{Tammann:93}
   Tammann, G.A. 1993,
   in: {\em Planetary Nebulae}, eds. R. Weinberger \&  A. Acker,
   IAU Symp. {\bf 155}, p.~515.
%
\bibitem[Tammann \& Kraan-Korteweg(1978)]{Tammann:Kraan-Korteweg:78}
   Tammann, G.A., \& Kraan-Korteweg, R. 1978,
   in: {\em Large Scale Structure of the Universe},
   eds. M.S. Longair \& J. Einasto, 
   IAU Symp. {\bf 79}, p.~71.
%
\bibitem[Tammann, Reindl, \& Thim(2001a)]{Tammann:etal:01a}
   Tammann, G.A., Reindl, B., \& Thim, F. 2001a,
   in: {\em Cosmology and Particle Physics}, eds. R. Durrer,
   J. Garcia-Bellido, \& M. Shaposhnikov, (Melville: Am. Inst. Phys.),
   p.~226. 
%
\bibitem[Tammann et~al.(2001b)]{Tammann:etal:01b}
   Tammann, G.A., Reindl, B., Thim, F., Saha, A., \& Sandage,
   A. 2001b,
   in: {\em A New Era in Cosmology}, eds. T. Shanks \& N. Metcalfe,
   ASP Conf. Ser., astro-ph/0112489.
%
\bibitem[Tammann \& Sandage(1968)]{Tammann:Sandage:68}
   Tammann, G.A., \& Sandage, A. 1968,
   {\em ApJ} {\bf 151}, 825.
%
\bibitem[Tammann \& Sandage(1985)]{Tammann:Sandage:85}
   Tammann, G.A., \& Sandage, A. 1985,
   {\em ApJ} {\bf 294}, 81.
%
\bibitem[Tammann \& Sandage(1995)]{Tammann:Sandage:95}
   Tammann, G.A., \& Sandage, A. 1995,
   {\em ApJ} {\bf 452}, 16.
%
\bibitem[Tammann \& Sandage(1999)]{Tammann:Sandage:99}
   Tammann, G.A., \& Sandage, A. 1999,
   in: {\em Harmonizing Cosmic Distance Scales in a Post-Hipparcos
   Era}, eds. D. Egret \& A. Heck, ASP Conf. Ser. {\bf 167}, p.~204.
%
\bibitem[Teerikorpi(1984)]{Teerikorpi:84}
   Teerikorpi, P. 1984,
   {\em A\&A} {\bf 141}, 407.
%
\bibitem[Teerikorpi(1987)]{Teerikorpi:87}
   Teerikorpi, P. 1987,
   {\em A\&A} {\bf 173}, 39.
%
\bibitem[Teerikorpi(1997)]{Teerikorpi:97}
   Teerikorpi, P. 1997,
   {\em ARA\&A} {\bf 35}, 101.
%
\bibitem[Teerikorpi et~al.(1999)]{Teerikorpi:etal:99}
   Teerikorpi, P., Ekholm, T., Hanski, M.O., \& Theureau, G. 1999,
   {\em A\&A} {\bf 343}, 713.
%
\bibitem[Theureau et~al.(1997)]{Theureau:etal:97}
   Theureau, G., Hanski, M., Ekholm, T., Bottinelli, L., Gouguenheim,
   L., Paturel, G., \& Teerikorpi, P. 1997,
   {\em A\&A} {\bf 322}, 730.
%
\bibitem[Theureau(2000)]{Theureau:00}
   Theureau, G. 2000,
   in: XIXth Texas Symposium, eds. E. Augbourg et~al., Mini-Symp. 13/12.
%
\bibitem[Tonry \& Schneider(1988)]{Tonry:Schneider:88}
   Tonry, J. \& Schneider, D.P. 1988,
   {\em AJ} {\bf 96}, 807.
%
\bibitem[Tripp(1998)]{Tripp:98}
   Tripp, R. 1998,
   {\em A\&A} {\bf 331}, 815.
%
\bibitem[Tully \& Fisher(1977)]{Tully:Fisher:77}
   Tully, R.B., \& Fisher, J.R. 1977,
   {\em A\&A} {\bf 54}, 661.
%
%
\bibitem[Udalski et~al.(1999a)]{Udalski:etal:99a}
   Udalski, A., et~al. 1999a,
   {\em AcA} {\bf 49}, 223.
%
\bibitem[Udalski et~al.(1999b)]{Udalski:etal:99b}
   Udalski, A., et~al. 1999b,
   {\em AcA} {\bf 49}, 437.
%
\bibitem[van den Bergh(1960)]{vandenBergh:60}
   van den Bergh, S. 1960,
   {\em J.~R. Ast. Soc. Canada} {\bf 54}, 49.
%
\bibitem[van den Bergh et~al.(1985)van den Bergh, Pritchet, \&
   Grillmair]{vandenBergh:etal:85} 
   van den Bergh, S., Pritchet, C., \& Grillmair, C. 1985,
   {\em AJ} {\bf 90}, 595.  
%
\bibitem[van Maanen(1923)]{vanMaanen:23}
   van Maanen, A. 1923,
   {\em ApJ} {\bf 57}, 264 (and references therein).
%
\bibitem[Wirtz(1918)]{Wirtz:18}
   Wirtz, C. 1918,
   {\em A.N.} {\bf 206}, 109.
%
\bibitem[Wirtz(1925)]{Wirtz:25}
   Wirtz, C. 1925,
   {\em Scientia} {\bf 38}, 303.
%
\bibitem[Yahil, Tammann, \& Sandage(1977)]{Yahil:etal:77}
   Yahil, A., Tammann, G.A., \& Sandage, A. 1977, 
   {\em ApJ} {\bf 217}, 903.
%
\bibitem[Zwicky(1933)]{Zwicky:33}
   Zwicky, F. 1933,
   {\em Helv. Phys. Acta} {\bf 6}, 110.

%
\end{thebibliography}
\end{document}